\RequirePackage{snapshot}
\documentclass[10pt,twocolumn]{z2-article}

\usepackage{framed}
\usepackage{float}
\usepackage{times}
\usepackage{oneinchmargins}
\usepackage{setspace}
\usepackage{authblk}
\usepackage{multirow}
\usepackage{amsmath}
\usepackage{centernot}
\usepackage{amssymb}
\usepackage[hyphens]{url}
\usepackage{alltt}
\usepackage{listings}
\usepackage{cite}
\usepackage{breakurl}
\usepackage{fancyvrb}
\usepackage{relsize}
\usepackage{soul}
\usepackage[normalem]{ulem}
\usepackage{array}
\usepackage[position=bottom]{subfig}
\usepackage{booktabs}
\usepackage{enumitem}
\usepackage{currfile}
\usepackage[usenames,dvipsnames,svgnames,table]{xcolor}
\usepackage{tikz}
\usepackage{pbox}
\usepackage{xstring}
\usepackage{xspace}
\usepackage{epstopdf}
\usepackage{graphicx}
\usepackage{float}
\usepackage{makecell}
\usepackage{url}
\usepackage{fixltx2e}
\usepackage{textalpha}
\usepackage{grffile}
\usepackage{titling}
\usepackage[pdftex,colorlinks,urlcolor=NavyBlue,linkcolor=NavyBlue,citecolor=NavyBlue,hyperfootnotes=false]{hyperref}

\makeatletter
\def\url@leostyle{%
  \@ifundefined{selectfont}{\def\UrlFont{\sf}}{\def\UrlFont{\small\ttfamily}}}
\makeatother
\floatstyle{boxed}
\newfloat{aside}{bth!}{lop}
\floatname{aside}{Aside}

\begin{document}
\include{preamble}
%\pagenumbering{gobble}

% Haryadi .. parskip, and par-indent
%\setlength\parindent{0pt}
%\setlength\parskip{5pt}

\newcommand{\beforesec}{\vspace{-.4cm}}
\newcommand{\aftersec}{\vspace{-.3cm}}

\newcommand{\beforesect}{\vspace{-.05cm}}
\newcommand{\aftersect}{\vspace{-.3cm}}

\newcommand{\beforesub}{\vspace{-.3cm}}
\newcommand{\aftersub}{\vspace{-.25cm}}

\newcommand{\beforesubsub}{\vspace{-.2cm}}
\newcommand{\aftersubsub}{\vspace{-.2cm}}

 \newcommand{\zsection}[1]{\section{#1}}
 \newcommand{\zsubsection}[1]{\subsection{#1}}
 \newcommand{\zsubsubsection}[1]{\subsubsection{#1}}

\newcommand{\smush}{0.25in}

\newcommand{\figWidthOne}{3.05in} 
\newcommand{\figWidthHalf}{1.45in} 
\newcommand{\figWidthTwo}{3.05in} 
\newcommand{\figWidthTwop}{1.6in} 
\newcommand{\figWidthThree}{2.2in} 
\newcommand{\figWidthSix}{1.1in} 
\newcommand{\figWidthFour}{1.7in} 
\newcommand{\figHeight}{2.0in}
\newcommand{\captionText}[2]{\textbf{#1} \textit{\small{#2}}}

\newcommand{\zbullet}{\hspace{-0.1cm}$\bullet$}

\newcommand{\eg}{\textit{e.g.}}
\newcommand{\ie}{\textit{i.e.}}
\newcommand{\etal}{\textit{et al.}}
\newcommand{\apriori}{\textit{a priori}}

%-------------------------------------------------------------------
% Haryadi -- new command
\newcommand{\msub}[1]{\vspace{1pt}\noindent{\bf #1}}

\newcommand{\ts}[1]{{\tt{\small#1}}}
\newcommand{\tsb}[1]{{\tt{\small{\bf#1}}}}
\newcommand{\tse}[1]{{\tt{\small{\em#1}}}}
\newcommand{\tss}[1]{{\tt{\footnotesize#1}}}
\newcommand{\exc}{$^{\ddag}$}        % except
\newcommand{\EIO}{\ts{EIO}}
\newcommand{\ENOSPC}{\ts{ENOSPC}}
\newcommand{\EDQUOT}{\ts{EDQUOT}}

%-------------------------------------------------------------------
% For fault injection table
\newcommand{\ip}{bad}
\newcommand{\nullp}{$\emptyset$}

\newcommand{\oops}{o}
\newcommand{\dead}{$\times$}
\newcommand{\alive}{$\surd$}
\newcommand{\unuse}{$\times$}
\newcommand{\use}{$\surd$}
\newcommand{\ic}{$\times$}
\newcommand{\con}{$\surd$}
\newcommand{\gpf}{G}
\newcommand{\npe}{null-pointer}
\newcommand{\usebuta}{$\surd^a$} % unmountable
\newcommand{\hwdetect}{d}

%% TABLE 2
\newcommand{\lateoops}{o$^b$}  % oops happens, but late
\newcommand{\lategpf}{G$^b$}   % gpf, but late
\newcommand{\iop}{i}           % invalid opcode
\newcommand{\detects}{d}       % assertion
\newcommand{\silentret}{s}     % app fails silently
\newcommand{\errorret}{e}      % app returns an error
\newcommand{\appworks}{$\surd$}     % app works!
\newcommand{\usebutar}{$\surd^{ar}$} % read-only and unmountable

\newcommand{\tbls}{\hspace{0.025in}}
\newcommand{\tblss}{\hspace{0.015in}}
\newcommand{\shrinkless}{\vspace{-0.01cm}}

%-------------------------------------------------------------------

%-------------------------------------------------------------------

% axes         
\newcommand{\x}{{\em x}}
\newcommand{\y}{{\em y}}
\newcommand{\xaxis}{x-axis}
\newcommand{\yaxis}{y-axis}

\newcommand{\KB}{~KB}
\newcommand{\KBs}{~KB/s}
\newcommand{\Kbs}{~Kbit/s}
\newcommand{\mbs}{~Mbit/s}
\newcommand{\MB}{~MB}
\newcommand{\GB}{~GB}
\newcommand{\MBs}{~MB/s}
\newcommand{\mus}{\mbox{$\mu s$}}
\newcommand{\ms}{\mbox{$ms$}}
\newcommand{\ns}{\mbox{$ns$}}

\newcommand{\unix}{{\sc Unix}}

\newcommand{\bquote}{\vspace{-0.25cm} \begin{quote}}
\newcommand{\equote}{\end{quote}\vspace{-0.05cm} }

\newcommand{\zquote}[2]{\begin{quote}
#1 --
{\em``#2'' }
%{\bf -- #1 }
\end{quote}}

\newcommand{\XXX}[1]{{\small {\bf (XXX: #1)}}}

\newcommand{\XXXX}{{\bf XXX}}
\newcommand{\xx}{{\bf XXX}}

% normal
\newcommand{\beforecaption}{\begin{spacing}{0.80}}
\newcommand{\aftercaption}{\end{spacing}}
\newcommand{\mycaption}[3]{{\beforecaption\caption{\label{#1}{\footnotesize \bf #2. } {\em \small #3}}\aftercaption}}

\newcommand{\sref}[1]{\S\ref{#1}}

\newcommand{\xxx}[1]{  \underline{ {\small {\bf (XXX: #1)}}}}

\newenvironment{packeditemize}{
\begin{itemize}
  \setlength{\itemsep}{1pt}
  \setlength{\parskip}{0pt}
  \setlength{\parsep}{0pt}
}{\end{itemize}}

\newcommand{\smalltt}[1]{\texttt{\fontsize{8.7}{5}\selectfont #1}}
\newcommand{\llename}{\smalltt{LLE}}
\newcommand{\mapname}{\smalltt{LLE-MAP}}
\newcommand{\mapnames}{\smalltt{LLE-MAPs}}
\newcommand{\numsys}{eight}
\newcommand{\numvuls}{26}
\newcommand{\fwrong}{$\times$}
\newcommand{\smallit}[1]{\textit{\scriptsize #1}}
\newcommand{\verysmalltt}[1]{\texttt{\scriptsize #1}}
\newcommand{\verysmall}[1]{\scriptsize #1}
\newcommand*\rot{\rotatebox{90}}
\newcommand{\hardraftname}{\mbox{\textsc{\fontsize{9.2}{5}\selectfont Haft}}}
\newcommand{\giname}{\mbox{\textsc{\fontsize{9.2}{5}\selectfont Saucr}}}
\newcommand{\ginamesmall}{\mbox{\textsc{\fontsize{7.2}{5}\selectfont Saucr}}}
\newcommand{\ginamebig}{\mbox{\textsc{\fontsize{10}{6}\selectfont Saucr}}}
\newcommand{\storesys}{\mbox{\textsc{\fontsize{10}{6}\selectfont Sacs}}}
\newcommand{\parname}{\mbox{\textsc{\fontsize{9.2}{5}\selectfont Par}}}
\newcommand{\ctrlname}{\mbox{\textsc{\fontsize{9.2}{5}\selectfont Ctrl}}}
\newcommand{\rasorname}{\mbox{\textsc{\fontsize{9.2}{5}\selectfont Rasor}}}
\newcommand{\sysname}{\mbox{\textsc{\fontsize{9.2}{5}\selectfont Sazerac}}}
\newcommand{\sysnamebig}{\mbox{\textsc{\fontsize{10}{6}\selectfont Sazerac}}}
\newcommand{\sysnamesmall}{\mbox{\textsc{\fontsize{7.2}{5}\selectfont Sazerac}}}
% ctrl local storage layer
\newcommand{\ctrlstore}{\mbox{\textsc{\fontsize{9.2}{5}\selectfont Clstore}}}
\newcommand{\ctrlstoresmall}{\mbox{\textsc{\fontsize{7.2}{5}\selectfont Clstore}}}

\newcommand{\ctrlnamesmall}{\mbox{\textsc{\fontsize{7.2}{5}\selectfont Ctrl}}}
\newcommand{\ctrlnaive}{\mbox{\textsc{\fontsize{9.2}{5}\selectfont Ctrl-Naive}}} 
\newcommand{\ctrlnaivesmall}{\mbox{\textsc{\fontsize{7.2}{5}\selectfont Ctrl-Naive}}} 

\newcommand{\ctrlraft}{\mbox{\textsc{\fontsize{9.2}{5}\selectfont Ctrl-Raft}}} 
\newcommand{\ctrlzab}{\mbox{\textsc{\fontsize{9.2}{5}\selectfont Ctrl-Zab}}} 
\newcommand{\errfsname}{\mbox{\textit{\fontsize{11.2}{5}\selectfont errfs}}}
\newcommand{\errbenchname}{\mbox{\textit{\fontsize{11.2}{5}\selectfont errbench}}}
\newcommand{\Kassandra}{Kassandra}
\newcommand{\microprogram}{microprogram}
\newcommand{\microprograms}{microprograms}
\newcommand{\microinstruction}{microinstruction}
\newcommand{\microinstructions}{microinstructions}
\newcommand{\Microinstructions}{Microinstructions}
\newcommand{\writeSC}{\smalltt{write()}}
\newcommand{\fsyncSC}{\smalltt{fsync()}}
\newcommand{\msyncSC}{\smalltt{msync()}}
\newcommand{\fdatasyncSC}{\smalltt{x fdatasync()}}
\newcommand{\linkSC}{\smalltt{link()}}
\newcommand{\mkdirSC}{\smalltt{mkdir()}}
\newcommand{\fempty}{$\phi$}
\newcommand{\fexists}{$\surd$}
\newcommand{\creatSC}{{\smalltt{creat()}}}
\newcommand{\unlinkSC}{{\smalltt{unlink()}}}
\newcommand{\renameSC}{{\smalltt{rename()}}}
\newcommand\floor[1]{\lfloor#1\rfloor}
\newcommand\ceil[1]{\lceil#1\rceil}
\newcommand{\totbugs}{60}
\newcommand{\totapps}{11}
\newcommand{\totappsw}{eleven}
\newcommand*{\combination}[2]{{}^{#1}C_{#2}}

\newcommand{\used}{$\surd$}
\newcommand{\usedpar}{$P$}
\newcommand{\useddollar}{$\surd$\textsuperscript{\$}}
\newcommand{\usedadler}{$\surd$\textsuperscript{a}}
\newcommand{\usedparstar}{$P$\textsuperscript{$*$}}
\newcommand{\notused}{}
\newcommand{\numapps}{eight}

\if 0 % colored version
\newcommand{\yes}{$\surd$}
\newcommand{\yesi}{\colorbox{gray!30}{$\surd$\textsubscript{$i$}}}
\newcommand{\nolow}{\colorbox{gray!30}{$\times$\textsubscript{$l$}}}
\newcommand{\no}{\colorbox{gray!85}{$\times$}}
\newcommand{\complower}{$L$}
\newcommand{\compmoder}{$M$}
\newcommand{\comphigher}{\colorbox{gray!85}{$H$}}
\newcommand{\na}{\footnotesize na}
\fi

\newcommand{\yes}{$\surd$}
\newcommand{\yesi}{$\surd$\textsubscript{$i$}}
\newcommand{\nolow}{$\times$\textsubscript{$l$}}
\newcommand{\nomod}{$\times$\textsubscript{$m$}}
\newcommand{\no}{$\times$}
\newcommand{\yeslow}{$\surd$\textsuperscript{$l$}}
\newcommand{\complower}{$L$}
\newcommand{\compmoder}{$M$}
\newcommand{\comphigher}{$H$}
\newcommand{\na}{\footnotesize na}

\newcommand*{\termindex}[2]{$\langle$\textit{epoch}:{#1}, \textit{index}:{#2}$\rangle$}
\newcommand*{\epochindex}[2]{{#1}.{#2}}
\newcommand*{\termindexnovar}{$\langle$\textit{epoch}, \textit{index}$\rangle$}
\newcommand*{\snapid}{$\langle$\textit{snap-index}, \textit{chunk}\#$\rangle$}
\newcommand*{\rafttermindexnovar}{$\langle$\textit{term}, \textit{index}$\rangle$}
\newcommand{\quotes}[1]{``#1''}
\newcommand{\camera}[1]{\textcolor{Black}{#1}}
\newcommand{\addcamera}[1]{\textcolor{Black}{#1}}
\newcommand{\todo}[1]{\textcolor{Red}{#1}}
%\newcommand{\camera}[1]{\textcolor{Black}{#1}}
%\newcommand{\addcamera}[1]{\textcolor{Black}{#1}}

%% OSDI 2025

\newcommand{\zc}[0]{Zerrow} % overall system
\newcommand{\da}[0]{KernelZero} % kernel module
\newcommand{\framework}{\da{}}

\newcommand{\blue}[1]{\textcolor{Blue}{#1}}

\begin{spacing}{1.0}
\setlength{\droptitle}{-1.4cm}
\date{}
\title{\Large \bf Zerrow: True Zero-Copy Arrow Pipelines in \textit{Bauplan} \vspace{-1.0em}}

\date{\vspace{-0.75em}\textit{}\vspace{-2em}}

% \affil{
%         \vspace{-0.2in}
%         {\normalsize\textit{University of Wisconsin -- Madison}}
%         {\vspace{-0.2in}}
%         }
\setlength{\droptitle}{-1.4cm}
\date{}
\title{\Large \bf Zerrow: True Zero-Copy Arrow Pipelines in \textit{Bauplan} \vspace{-1.0em}}
\author{Yifan Dai$^\ast$, Jacopo Tagliabue$^\star$, Andrea Arpaci-Dusseau$^\ast$, \\ Remzi Arpaci-Dusseau$^\ast$, Tyler R. Caraza-Harter$^{\ast\star}$
}
\date{\vspace{-0.75em}\textit{$^\ast$ University of Wisconsin--Madison}, \textit{$^\star$ Bauplan Labs}\vspace{-0.9em}}
\maketitle

\noindent
\textbf{\textit{Abstract.}} Bauplan is a FaaS-based lakehouse specifically built for data pipelines: its  execution engine uses Apache Arrow for data passing between the nodes in the DAG. While Arrow is known as the ``zero copy format'', in practice, limited Linux kernel support for shared memory makes it difficult to avoid copying entirely. In \textit{this} work, we introduce several new techniques to eliminate nearly all copying from pipelines: in particular, we implement a new kernel module that performs de-anonymization, thus eliminating a copy to intermediate data. We conclude by sharing our preliminary evaluation on different workloads types, as well as discussing our plan for future improvements.

\beforesect
\section{Introduction}
\label{sec:intro}
\beforesect

Data pipelines are a popular programming paradigm for data analysis and machine-learning workloads.  Data pipelines are frequently implemented as DAGs (Directed Acyclic Graphs), where each node of the DAG describes a transformation to perform on the data~\cite{Zaharia-12-Spark, Apache-Airflow, Luigi, Tagliabue2023ReasonableSM}. Given fine-grained nodes, efficient communication between nodes is especially important for good performance~\cite{Jia-21-Nightcore,Xin2018HowDI,shankar-operationalizing-ml}. Much work has addressed how to efficiently communicate between DAG nodes in a distributed setting; for example,  using compact representations with variable-length encoding and compression to minimize network I/O, and key-based partitioning and shuffling to group related records on the same machine~\cite{Dean-04-MapReduce,luan-exoshuffle,Zaharia-12-Spark,Perron-20-Starling}. 

\textit{Bauplan} is a lakehouse built in the spirit of ``composable data systems'' \cite{Tagliabue2023BuildingAS,10.14778/3603581.3603604}. While presenting to users a unified API for assets and compute, pipelines, and queries \cite{10.1145/3650203.3663335}, the underlying system is built entirely out of containerized, ephemeral functions over object storage; \ie{},
data pipelines run as a chain of functions spawn in milliseconds inside off-the-shelf VMs  \cite{10.1145/3702634.3702955}. Subscribing to the  ``Reasonable Scale'' hypothesis~\cite{10.1145/3460231.3474604,McSherry2015ScalabilityBA}, Bauplan targets single node transformations: while memory is becoming cheaper (\eg{}, from \$4K for 1~TB of RAM in 2014 to \$1K in 2023~\cite{worlddata}), data sets are not that much larger (\eg{}, ``Big Data era''~\cite{doi:10.1089/big.2013.0037} vs.\ today's Machine Learning datasets~\cite{CoveoSIGIR2021}). The intermediate format for data passing is Apache Arrow~\cite{Apache-Arrow,arrow-topol}, a popular memory layout for in-memory and over-the-wire tabular representation.

Although people regard Apache Arrow~as a ``zero copy'' format for in-memory data, Arrow only removes data copying on the reading node's side. To eliminate reader-side copies, Arrow avoids pointers so that data can be mapped to different locations in different address spaces; to reduce memory duplication, the same intermediate data can also be mapped by multiple downstream nodes. Unfortunately, simply using Arrow for inter-node communication does not eliminate several sources of copying and duplication in data pipelines.  First, many tools and libraries that return Arrow data allocate space with malloc, which uses anonymously mapped memory without a backing file; operating systems (including Linux) do not typically support sharing of anonymous memory, so unless all libraries in the Arrow ecosystem are rewritten to use shared memory, a copy to shared memory is necessary.  Second, DAG nodes must perform copies when Arrow output overlaps with Arrow input (\eg{}, the node adds a column to an input table), as the existing Arrow IPC protocol does not provide a way to identify or reference such overlap. Finally, when independent DAGs deserialize the same data from on-disk formats (\eg{}, Parquet files) to Arrow, different processes contain identical copies of Arrow data.

In pursuit of ``true zero-copy'' in a data DAG, we introduce \zc{}, an experimental system focused on i) our new kernel support for sharing anonymous memory and ii) our new extended Arrow IPC protocol. \zc{} introduces several new techniques.  First, \zc\ introduces \textit{de-anonymization}; our new kernel module (\da) enables Linux processes to transfer ownership of anonymous memory regions to in-memory files that can be mapped into other address spaces. Second, \zc\ extends the Arrow IPC protocol to reference these in-memory files instead of copying the data; \zc\  implements \textit{IPC inspection} to identify cases where outputs overlap with inputs, and then applies \textit{resharing} to refer back to the input data instead of copying the actual data into the output files. Third, to avoid duplication of deserialized Arrow data in different DAGs, \zc\ provides a centralized \textit{shared deserialization} service: if multiple DAGs use the same input data within a short-enough time interval, each DAG will reference a single physical copy of the data.  

\zc's techniques provide substantial performance gains in a variety of scenarios by reducing both the time overhead of copying identical data and the memory overhead of duplicating identical data; this memory reduction allows significantly more DAG nodes to run in parallel. De-anonymization eliminates write-side copies, halving the latency of a single node outputting Arrow data that was loaded from a Parquet file. For a large batch of 2-node DAGs loading data from the same source, shared deserialization improves throughput by 2.8x.  IPC inspection and resharing dramatically reduce physical output sizes for a variety of transformations, in some cases to practically zero new data.  In combination, these features improve overall throughput by 1.5-9.5x for a complex mix of jobs, depending on parallelism and DAG depth.

The rest of this paper is organized as follows.  We give more background regarding Arrow and where copying and duplication often occur (\S\ref{sec:bg}).  We then introduce \zc{}'s design (\S\ref{sec:design2}) and implementation on Linux (\S\ref{sec:implementation}).  Finally, we evaluate \zc{}'s performance (\S\ref{sec:eval}), describe related work (\S\ref{sec:related}), and conclude (\S\ref{sec:conc}).
\beforesect
\section{Background and Motivation}
\label{sec:bg}
\beforesect

\subsection{Background: Arrow-Based DAGs}

Our goal is to build a prototype that increases the memory-efficiency for workloads currently executed by the FaaS lakehouse Bauplan \cite{10.1145/3702634.3702955}. Our strategy is to implement new kernel-level support for shared memory and design every subsystem of our new platform around this support, thus avoiding the copying and duplication necessary in the original implementation due its reliance on the generic Linux kernel, which has limited sharing support.

Our workload consists of directed-acyclic graphs (DAGs), where each node transforms the data in some way.  Transformations run as fully independent containerized processes (Python or SQL functions) in order to maximize flexibility and avoid dependency collisions. Transformations have as input and output dataframe-like objects, which are persisted on disk (object storage) as compressed Parquet files, and shared in-memory as Arrow tables: Arrow data is uncompressed, immutable, and formatted such that computation can be performed on data in-place (often with SIMD instructions).

Arrow's computational format specifies that columns of fixed-length values (\eg{}, float64s) should be stored in contiguous, indexable arrays.  Variable-length values such as byte or character strings and nested types (\eg{}, lists of floats) are also stored contiguously in memory in a value buffer.  For example, the first byte of a string at index N+1 in an array immediately follows the last byte of a string at index N.  An array of offsets (into the first buffer) are stored in a separate buffer, providing indexing over contiguous variable-length values.  Arrow's use of offsets rather than pointers requires an alternative representation for NULL values; Arrow uses ``null'' bitmaps for this purpose.  Sometimes, long strings or byte sequences may appear repeatedly in a column; Arrow supports an optional dictionary encoding where each unique value is stored once in a dictionary, and an array of numeric codes is used to reference specific dictionary entries.

Arrow also specifies an Arrow IPC (Inter-Process Communication) wire format for sending batches of records to a destination (\eg{}, 
a socket, file, or memory buffer).  A single schema message defines the columns and types for the entire stream and any number of batch messages may be written (metadata for indexing over batches may optionally be included).  Record batches use the same layout as Arrow data in the computational format.  If dictionary encoding is used for some of the columns, dictionary messages may be sent as well.  Our design (\S\ref{sec:design2}) exploits several aspects of these formats (contiguous layouts, pointer avoidance, dictionary encoding, etc.) to avoid copying and duplication.

\beforesect
\subsection{Problem Statement}
\label{sec:problem-statement}

\begin{figure}
    \centering
    \includegraphics[width=\linewidth]{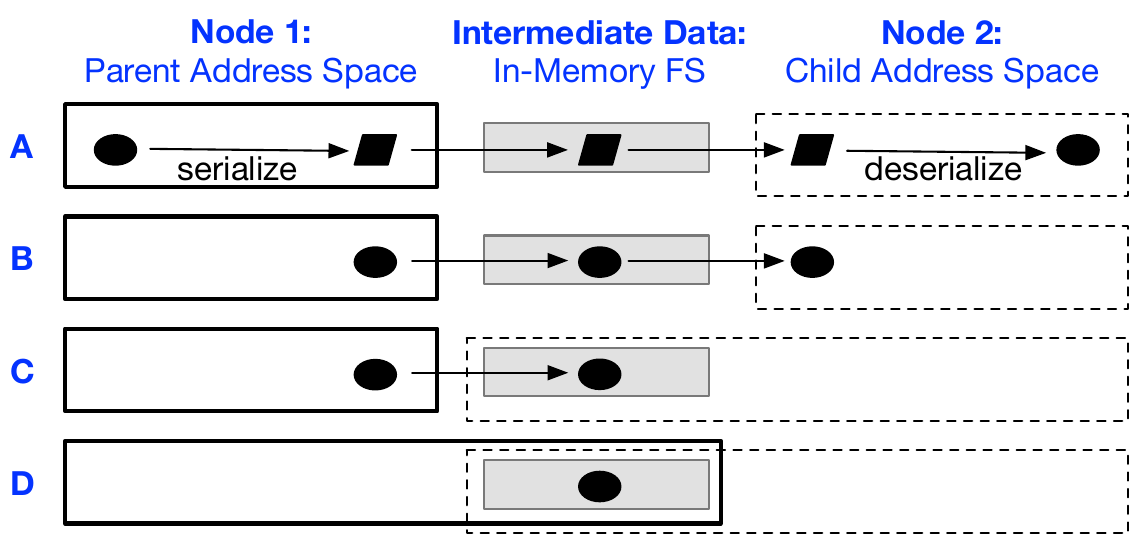}
    \caption{\textbf{Communication: Degrees of Zero Copy}} 
    \label{fig:zero-copy}
\end{figure}

Arrow is a so-called ``zero-copy'' format, but Figure~\ref{fig:zero-copy} illustrates that there are different degrees to which communication techniques avoid copies.  Example A illustrates a generic communication protocol that (unlike Arrow) does not attempt to avoid copies.  Copying between formats occurs during serialization and deserialization because the ``wire'' format differs from the ``computational'' format.  Wire formats often differ from corresponding computational formats that are based on pointers (which are not valid in other address spaces).  Wire formats are also often designed to represent data compactly (to save network I/O), but CPUs cannot directly compute over data in the compact format.  Example A also illustrates copies of the wire-formatted data, to and from an intermediate medium.

Example B shows that serialization and deserialization copies can be avoided when the computational and wire formats are the same.  This is the case with Arrow: a receiving process can obtain a tabular view over the record batch data in the Arrow IPC stream, without copying the data elsewhere (assuming matching endianness).  Example C shows further copy reduction: if the intermediate data is in an in-memory file system, the receiving process can simply map that data into its address space, without a reader-side copy.

Of course, Example D represents the ideal scenario. 
There is no write-side copy, and the reader, writer, and intermediate data simply reference the same physical data.  Unfortunately, this scenario is difficult to realize as most Arrow-compatible tools are not designed to directly populate shared memory, and most operating systems (in particular, Linux), do not already provide a mechanism to transfer ownership of a process's anonymous memory to an in-memory file.

\begin{figure}
    \centering
    \includegraphics[width=\linewidth]{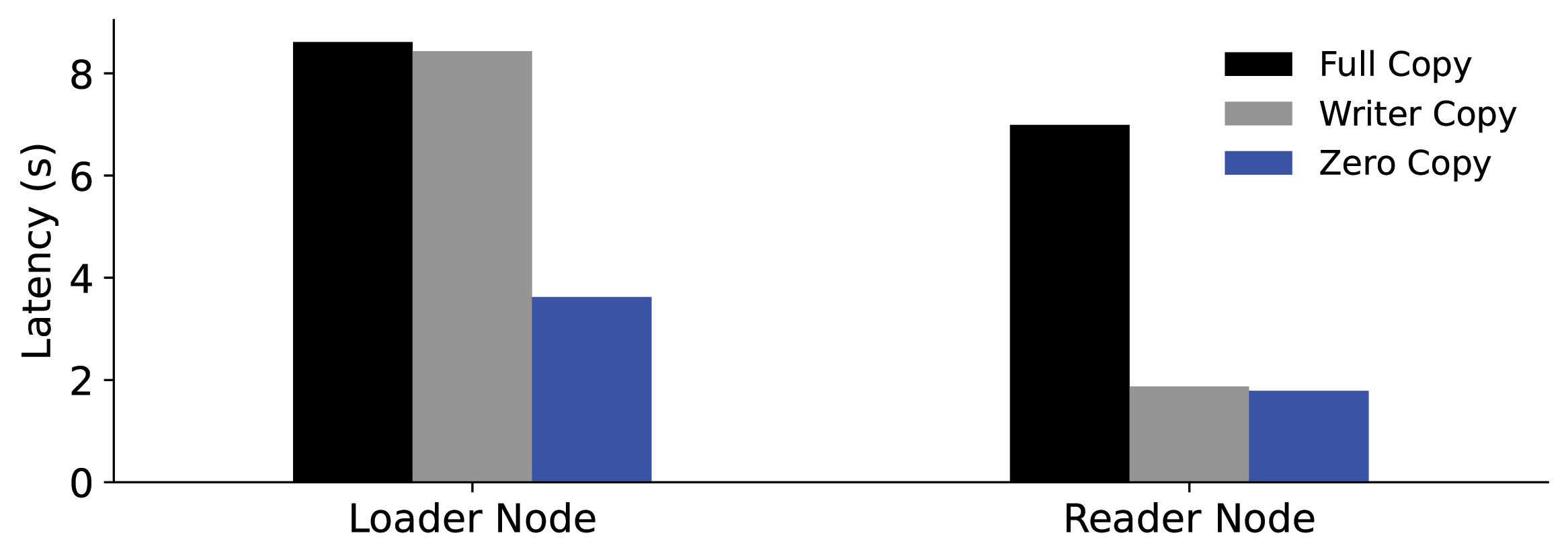}
    \vspace{-0.6cm}
    \caption{\textbf{Latency with Copy Avoidance}}    
    \label{fig:motivation-exp}
\end{figure}

Figure~\ref{fig:motivation-exp} gives a preview of the performance benefits that are possible when we implement such kernel support and modify the Arrow IPC protocol to leverage the new functionality.  For this experiment, we run a simple DAG with 2 nodes. The first node (Loader Node) loads a 10~GB Arrow table of 10 columns of integers from a Parquet file and outputs the Arrow IPC format. The second node (Reader Node) reads the Arrow IPC file and computes the sum of all integers in the table.  The three bars, Full Copy, Writer Copy, and Zero Copy correspond to approaches B, C, and D in Figure~\ref{fig:zero-copy}, respectively.  We see that Writer Copy and Zero Copy are 3.8$\times$ and 3.9$\times$ faster than Full Copy in Reader Node, and Zero Copy further performs 2.3$\times$ faster than Writer Copy in Loader Node, demonstrating the benefit of real zero copy.  Although deserializing a Parquet file to Arrow is of course more expensive than a simple memory copy, the former is a parallel operation whereas writes to the Arrow IPC format are performed serially.  We describe and evaluate our approach to eliminating write-side copies in more detail in subsequent sections (\S\ref{sec:design2}, \S\ref{sec:implementation}, and \S\ref{sec:eval}).

% \begin{figure}
%     \centering
%     \includegraphics[width=\linewidth]{figs/share_type.pdf}
%     % \vspace{-0.6cm}
%     \caption{\textbf{Sharing Opportunities.}
%     Rounded boxes represents in-memory data. Those with the same color are with the same data. Those with the same number are the same piece of physical data. Solid arrows indicate copying.}
%     \label{fig:share-type}
% \end{figure}

\beforesect
\subsection{Goals}
\label{sec:goals}

\zc\ has several design goals, related to supporting high-throughput execution for pipeline jobs (Goals 1-3) and a flexible developer experience (Goals 4-6).

\textbf{G1) Avoid duplicating data.}  Memory capacity is a valuable resource, and duplication wastes it.  Duplication may occur in deserialized data sharing if multiple downstream nodes in a DAG have their own copies of the same input, or when multiple DAGs deserialize Arrow data from the same source.

\textbf{G2) Avoid copying data between and within nodes.}  As we have shown (\S\ref{sec:bg}), copying \textit{between} nodes wastes memory bandwidth and results in slower executions.

\textbf{G3) Perform share-aware caching, scheduling, and eviction actions.}  Nodes will frequently be using data produced by other nodes.

\textbf{G4) Allow developers to write arbitrary node code}.  Many data pipeline tools are opinionated about how transformations can be represented.  For example, Spark SQL represents transformations as well-defined RDDs (this is conducive to a variety of query optimizations).  In contrast, \zc{} prioritizes flexibility in order to support the broad range of tasks data scientists and machine-leaning engineers perform (\eg{}, one may wish to implement a machine-learning algorithm other than those in Spark's MLlib).

\textbf{G5) Support a broad ecosystem of unmodified tools and libraries.}  Existing libraries rarely have specialized support for directly populating shared memory with Arrow data, and it is unrealistic to require existing libraries in the PyPI repo (or elsewhere) to be rewritten for our platform.

\textbf{G6) Support library version independence between nodes in a DAG.}  Version independence has multiple benefits.  One is that different downstream nodes with a common parent may deploy different versions of the same package (useful for identifying regression bugs or comparing accuracy across different model versions).  For complicated DAGs, version independence also solves issues related to version incompatibility for indirect dependencies.

\beforesect
\subsection{Challenges}
\label{sec:challenges}

Achieving our goals (\S\ref{sec:goals}) in combination is non-trivial.  Copy and duplication avoidance (G1 and G2) require \zc\ to pervasively use shared memory, which brings new challenges to resource management and scheduling (G3). Supporting a wide range of unmodified code (G4 and G5) with library version independence (G6) require isolated execution of unmodified user code and library. We summarize them into the following challenges regarding independent code execution, copy avoidance, and resource management:

\textbf{Challenge 1:} \textit{How can \zc\ enable each node in a DAG to share memory while having distinct dependencies (G1, G2, and G6)?}  Sharing (required to avoid copying and duplication) would be easier if nodes could share a single address space, but supporting different dependency versions requires nodes to run in distinct processes.

\textbf{Challenge 2:} \textit{How can \zc\ avoid having duplicate data in memory when different DAGs start by deserializing the same data (G1)?}  Different DAGs are written by different developers and may not be submitted at the same moment, so platform-level support is needed to avoid duplication.

\textbf{Challenge 3:} \textit{What mechanisms can \zc\ provide to free the physical memory that backs virtual Arrow artifacts (G3)?}  When memory is freed by swapping Arrow data to disk, we do not have direct control over which data is swapped; \zc\ will need to identify a way to influence which intermediate data is evicted.

\textbf{Challenge 4:} \textit{How can nodes avoid copies when producing intermediate data (G2, G5)?} Arrow used in combination with file mappings makes it easy to eliminate copies on the read side, but existing kernel support does not allow a process to expose a range of anonymous memory to other processes for mapping, and we may not modify existing Arrow-based libraries to directly populate shared memory.

\textbf{Challenge 5:} \textit{How can we avoid copying within a node when there is redundancy between inputs and outputs (G2)?}  This is challenging because Arrow IPC does not provide a way to reference regions of inputs (without copying those regions).  A further complication is that some operations (\eg{}, sorting and filtering) may result in overlap at row granularity.

\textbf{Challenge 6:} \textit{How can we expose details regarding sharing between different virtual Arrow artifacts to caching policy logic (G3, G4, G5)?}  Arrow artifacts may be produced by arbitrary user code and libraries, so we do not have the opportunity to trace through operations to identify overlap.

\beforesect
\section{\zc{} Design}
\label{sec:design2}
\beforesect

We introduce \zc{}, a new data pipeline platform that uses shared Arrow data to efficiently pass intermediate data with minimal copying and duplication.  \zc{} uses a mix of existing techniques (\textit{share mounting} and \textit{limit dropping}) and novel techniques (\textit{shared deserialization}, \textit{de-anonymization}, \textit{resharing}, \textit{dictionary sharing}, and \textit{IPC inspection}) to overcome the challenges we have described (\S\ref{sec:challenges}).  We provide an overview of \zc{}'s major components, (\S\ref{sec:arch}), trace through the zero-copy write path (\S\ref{sec:zero-copy-design}), and describe \zc{}'s memory management policies (\S\ref{sec:policy}).

\beforesect
\subsection{\zc{} Overview}
\label{sec:arch}

\begin{figure}[tp]
    \centering
    \includegraphics[width=\linewidth]{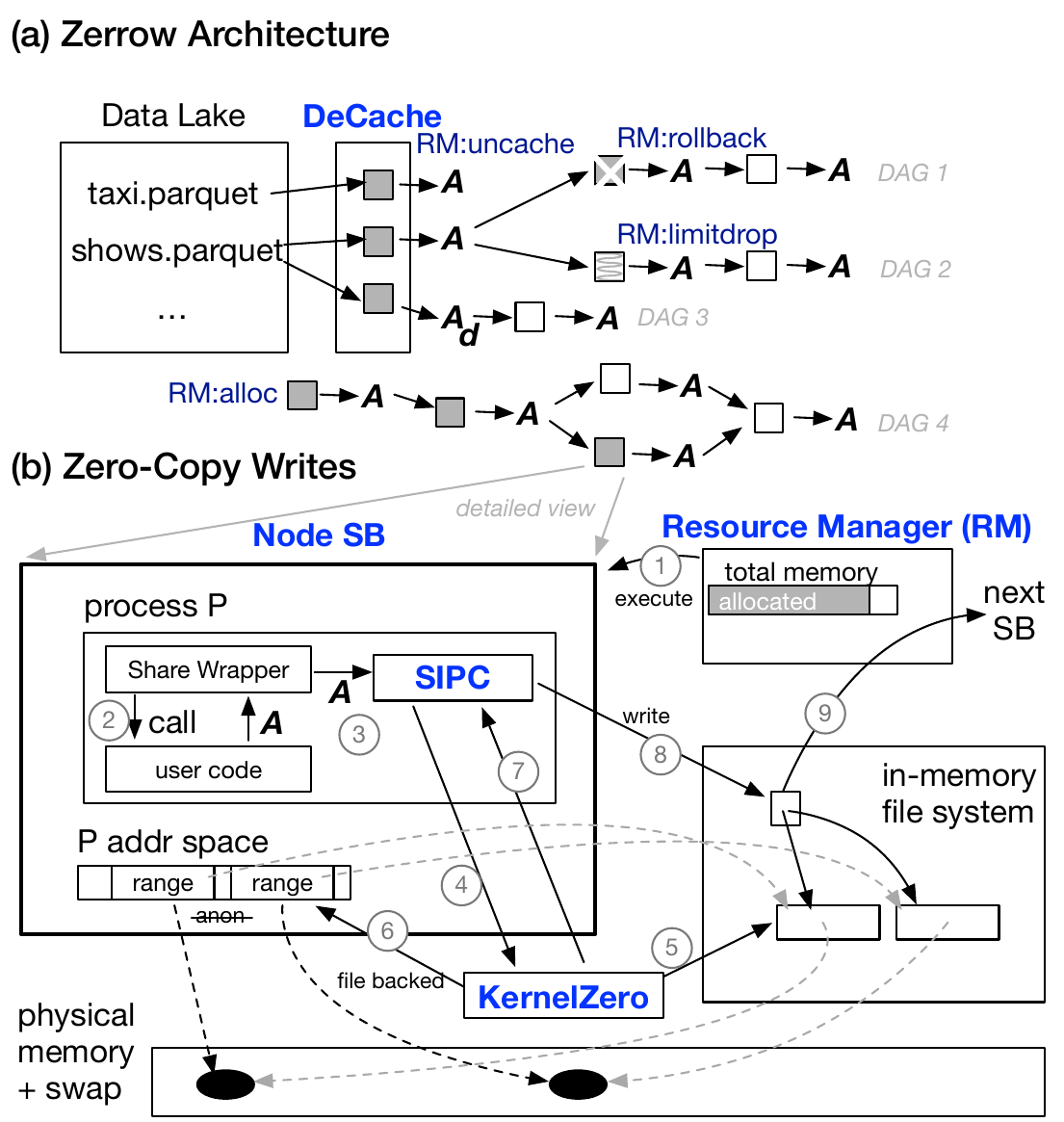}
    \caption{\zc{} Architecture and Write Path}
    \label{fig:arch}
\end{figure}

Figure~\ref{fig:arch}a illustrates \zc{}'s architecture and four running DAGs (1-4), most of which are using Parquet files as a source.  Each square represents a node in a DAG, and shading indicates a node is running or completed.  Intermediate Arrow data (``\textit{A}s'') is passed via shared memory between nodes.

In order to achieve data sharing and independence between nodes in terms of code resources (\textbf{Challenge 1}), each process runs in its own container, called a \textit{node sandbox}, and all the containers mount a common in-memory file system for sharing data.  This \textit{share mounting} technique is a standard containerization feature (\eg{}, a Docker {\tt --ipc=host} option provides such a mount).

DAG nodes can either provide custom code or indicate that generic loader code should deserialize specified Parquet data to Arrow.  When multiple DAGs rely on the same inputs, multiple nodes may map the same deserialized inputs (generated by loader nodes) into their address spaces.  This \textit{deserialization sharing} avoids duplication of in-memory data across different DAGs (\textbf{Challenge 2}).  Loader nodes and their outputs are managed as a deserialization cache (DeCache) so that DAGs starting at different times can still use the same inputs.  Regular node sandboxes (and their outputs) are normally freed upon DAG completion, but DeCache sandboxes are preserved until memory pressure requires eviction.  Different DAGs may wish to deserialize the same data with or without dictionary encoding on certain columns.  In this case, different loader nodes will be used to create each representation (\eg{}, shows.parquet is deserialized with and without dictionary encoding in Figure~\ref{fig:arch}A).

A Resource Manager (RM) manages memory resources, including DeCache contents.  Figure~\ref{fig:arch}A shows four actions (labeled with the ``RM:'' prefix) that the RM can perform.  First, the RM decides when memory should be allocated to start running new sandboxes (\textit{RM:alloc}) (the RM acts as a scheduler, deciding which node should receive the allocation).  Second, the RM can perform three types of eviction action (\textit{RM:uncache}, \textit{RM:rollback}, and \textit{RM:limitdrop}) to free up memory as needed (\textbf{Challenge 3}).  Uncaching DeCache entries and rolling back DAG progress both involve deleting completed sandboxes and their Arrow outputs.  In the latter case, the node will need to be re-executed later so that the DAG can eventually complete.

An alternative to deleting sandboxes is to rely on kernel swap; however, we want to avoid swapping out intermediate data that will be consumed by nodes that will soon be scheduled to run.  Influencing swap is challenging due to limited kernel support for specifying what data in particular should be swapped out.  We borrow a technique, \textit{limit dropping}, from an early version of Senpai~\cite{meta-memory-offloading}.  With this approach, a container's memory limit is dropped to trigger swap.  In our case, in-memory files produced by a container are counted towards the container's memory limit, so the RM can select specific intermediate Arrow data to swap to disk.  In order to retain control over intermediate data after a node finishes running, we allow the process in the container to exit upon node completion, but retain the container itself until the whole DAG completes.

The eviction actions described operate on virtual Arrow artifacts, which may partially overlap with other artifacts in physical memory.  We will shortly describe how sharing details are exposed to the RM (\S\ref{sec:zero-copy-design}) and what RM policies govern the use of the mechanisms just described (\S\ref{sec:policy}).

\beforesect
\subsection{Zero-Copy Write Path}
\label{sec:zero-copy-design}

As shown, Arrow's use of a common format for wire data and computational data eliminates the need for serialization and deserialization copies, and reader-side memory mappings eliminate a reader-side copy from intermediate data (\S\ref{sec:problem-statement}).  We now describe our techniques for avoiding a writer-side copy and input-to-output copies.

Figure~\ref{fig:arch}b illustrates \zc{}'s write path: (1) the RM allocates memory to start a new sandbox, passing in references to input data, if any.  (2) Inside the sandbox, a \textit{share wrapper} invokes the provided user code.  That code returns Arrow data back to the wrapper.  The Arrow data may have been generated by any package using any Arrow implementation, so the data is unlikely to already reside in shared memory.  Instead, the Arrow data will usually reside in \textit{anonymous} memory (\ie{}, memory not backed by a file).  This is unfortunate, as processes in downstream nodes will need a reference to a file that can be mapped into their address spaces.

(3) The share wrapper sends a reference to the Arrow data to a \textit{SIPC} (Shared IPC) module, which extends the standard Arrow IPC implementation.  SIPC (like the unmodified version) is responsible for writing schemas, records, and dictionaries (as needed) to a sink, in this case an in-memory file.  SIPC will copy the schema to the sink as usual, but it will leverage new kernel support to share records and dictionaries without copying.

(4) SIPC sends the anonymous ranges of memory to a new kernel module, \da{}, that converts the anonymous memory to file-backed memory.  (5) \da{} allocates new in-memory files and updates metadata in the process's address space to indicate that the specified range corresponds to a section of the new file. (6) \da{} transfers ownership of previously anonymous memory to new in-memory files.  This transfer does not involve copying actual data, and afterwards both the process's address space and the in-memory file will point to the same physical memory pages.

(7) \da{} returns identifiers for the new files and offsets corresponding to the de-anonymized ranges.  (8) SIPC writes references to the new de-anonymized files to the SIPC sink.  A SIPC file will be much smaller that an Arrow IPC file because it has references to the data instead of the actual data.  This \textit{de-anonymization} technique enables \zc{} to avoid copying data on the write path (\textbf{Challenge 4}).

For nodes in a non-trivial DAG, it will be common to have processes that are both consumers and producers of Arrow data, and some operations may produce overlap between the input and output.  Examples include selecting a subset of columns, slicing rows, concatenating tables (along either axis).  When SIPC reads inputs, it records the locations where it maps input data into the address space.  Upon later writes, SIPC recognizes addresses corresponding to input ranges.  This enables \textit{resharing}: SIPC writes references to input files to the output stream instead of copying the data within the node (\textbf{Challenge 5})

Resharing is sometimes possible with filtering and sorting too, but this is more challenging because inputs and outputs will overlap at fine-granularity (\ie{}, rows).  For regular encodings, \zc{} will unfortunately need to copy the data.  For dictionary encodings, however, SIPC uses a \textit{dictionary sharing} technique; references to input dictionaries are written to outputs, and only columns of dictionary lookup codes must be copied.  Consequently, dictionary encoding will often be useful for \zc{} even if there are no repeated strings in a column (the standard use case for dictionary encoding).

With resharing, intermediate data is represented as virtual Arrow tables referencing physical memory.  Reference details are very relevant to caching decisions (\eg{}, it may be necessary to delete outputs from multiple sandboxes to free up a physical memory with multiple references to it).  We cannot control the user code generating outputs, but SIPC inspects outgoing Arrow data and collects sharing information on behalf of the RM.  This \textit{IPC inspection} technique allows the RM to make informed eviction decisions regarding eviction and garbage collection (\textbf{Challenge 6}).

\beforesect
\subsection{Admission and Eviction Policy}
\label{sec:policy}

The RM uses a straightforward admission control that allows new sandboxes to be created when the available memory (not reserved by existing sandboxes) can satisfy the memory requirement of a node.  When multiple nodes are waiting for memory allocations, the RM assigns priority in depth-first order (\ie{}, the closest to finishing a DAG), as completing DAGs that are nearly finished enables deletion of all intermediate data for that DAG, freeing resources for other DAGs.

Admission control has the potential to cause deadlock if intermediate data has exhausted all system memory and all DAGs are waiting for memory to make progress; the intermediate data may not be releasable until DAG completion.  Even without deadlock concerns, too much intermediate data in memory could reduce the opportunity for more parallelism in DAG execution.  Kernel swapping is one solution, though excessive swapping can harm performance.  Thus, the RM uses DAG rollback and limit dropping (\S\ref{sec:zero-copy-design}) and knowledge of sharing details to free up memory as necessary.

The RM chooses between memory-freeing actions as follows.  First, the RM uncaches DeCache entries with no active references, if any. Then, the RM evicts the outputs of nodes with the lowest priority (as described based on depth).  Outputs are evicted one by one until the available memory is larger than the memory requirement of the node scheduled to run next. The RM has two mechanisms for evicting a node's output (\S\ref{sec:zero-copy-design}): \textit{rollback} deletes outputs of a function and rollback the pipeline, and \textit{limit dropping} swaps the outputs of a function to the swap space by manipulating container memory limits. The cost of deleting outputs depends on the latency of the function that generates the outputs and the cost of swapping depends on the size of the outputs. Thus, the RM adopts \textit{adaptive eviction}, which adaptively deletes or swaps the outputs of a function based on the ratio of its latency to the output sizes.  The threshold should be tuned offline and should depend on throughput of the swap device.

\if 0
Deadlock happens when the memory is filled up with intermediate in-memory data, which are the outputs of intermediate functions, and there is not enough memory to run more functions and make progress. One can rely on kernel swapping to make space, but kernel swapping is very slow because it may swap out data that are being used by running nodes or nodes about to be scheduled. Even when there is no deadlock, too many intermediate data in memory could reduce the opportunity for more parallelism in DAG execution. Thus, with our knowledge of references to the intermediate data, we develop eviction mechanisms of temporarily-unused intermediate data to resolve deadlocks and make space for more parallelism.

with our knowledge of references to the intermediate data, we develop eviction mechanisms of temporarily-unused intermediate data to resolve deadlocks and make space for more parallelism.
\fi

\beforesect
\section{Implementation: \zc{} on Linux}
\label{sec:implementation}
\beforesect

We now describe the implementation of \zc{}. We provide relevant background regarding Linux memory management (\S\ref{sec:linux-background}), then describe implementation of each \zc{} subsystem (\S\ref{sec:subsystem-implementations}).

\beforesect
\subsection{Background: Linux Memory}
\label{sec:linux-background}

On Linux, every process has its own virtual address space, which consists of a \textit{page table} and a set of \textit{virtual memory areas} (VMAs). Page tables support the translation of virtual addresses to physical addresses, at the granularity of pages (usually 4~KB), and VMAs describe various properties for contiguous ranges of virtual pages, such as permissions and backing device.  Calls to the {\tt mmap} (memory map) system call create new VMAs.  An {\tt mmap} call can map a file's contents as a new VMA into the address space, or a range of anonymous memory (used for heap and stack memory).  If a parent process forks a child, both will have virtual pages within anonymous VMAs pointing to the same physical memory.  Aside from fork, unrelated processes will not otherwise have different anonymous VMAs mapping to the same physical data.

These details have several implications for \da\: (1) memory for Arrow data is typically allocated from anonymous VMAs with the help of a malloc implementation. (2) DAGs are more complicated structures that process trees, so process fork is insufficient for sharing this anonymous data between processes running node code. (3) Any sharing of the same data within multiple address spaces will be at page granularity (usually 4~KB).  (4) The file-backed vs. anonymous distinction is at VMA granularity, a VMA used by the heap may be larger than data to be shared.

In addition to the limited sharing Linux provides via fork, Linux supports a POSIX API for shared memory.  POSIX shared memory integrates closely with tmpfs, an in-memory Linux file system; in particular, shared memory objects are represented as files in a tmpfs instance typically mounted at /dev/shm. Sharing is achieved with multiple processes use {\tt mmap} to created file-backed VMAs in their address spaces that refer to the same tmpfs file.

Note that in addition to the anonymous~vs.~file-backed distinction, {\tt mmap} orthogonally specifies whether a VMA should be {\tt MAP\_PRIVATE} (handle updates to memory in the VMA with copy-on-write) or {\tt MAP\_SHARED} (updates in one process should be visible externally).  We are only concerned with sharing for performance purposes.  Update visibility is irrelevant because Arrow data is immutable. Furthermore, a parent node completes before its children begin, and children map shared data using read-only permissions.

On Linux, memory pressure may be handled by swapping out pages to a swapfile residing on a disk-backed file system.  Both anonymous VMAs and tmpfs files can take advantage of swap.  When a page of anonymous memory is swapped out, a swap entry replaces the entry in the page table that previously referred to physical memory.  A tmpfs file has an array of entries that normally point to physical memory; much as with a page table, these tmpfs entries can be become ``swap entries'' when data is evicted.  De-anonymization must account for swap entries in page tables.

Space in physical memory and swap are both finite resources. Linux Control groups, commonly called \textit{cgroups}, are an abstraction that can be used to provide limits (for memory and other resources) to groups of processes.  Measuring memory consumption for a cgroup is non-intuitive when processes in different cgroups map the same files into their address spaces.  For files residing in regular, on-disk file systems, pages are charged to the cgroup whose process first loads them into memory (the page cache). Later processes mapping the same data are not charged, unless the data is evicted from memory and re-loaded by a different process.  Files in tmpfs are handled differently.  A cgroup ``owns'' the tmpfs files created by processes in the cgroup and space is the file is charged as memory consumption.  The original cgroup is discharged upon swapout, then charged again upon later swapin (assuming it still exists), even if a process in a different cgroup mapped the file and triggered the swapin.

Control groups are tightly integrated with swap.  Each cgroup has its own set of LRU queues for anonymous and file-backed pages, and under memory pressure, physical pages from the start of the queues are evicted from memory.

\subsection{Subsystems}
\label{sec:subsystem-implementations}

\subsubsection{\da\: File-Mapping Anonymous Memory}
\label{sec:deanon}

We implement the \da\ (de-anonymizer) subsystem as a new Linux module, written in C.  \da\ exposes two interfaces: \texttt{new\_file()} allocates a tmpfs file to receive anonymous memory in the future, and \texttt{deanon(file\_id, addr, len)} moves the anonymous memory in \texttt{[addr, addr+len)} to the end of the indicated tmpfs file (the operation resembles an append, without a copy).

A {\tt deanon} call traverses the page table to identify pages in the specified range. For each page, \da\ modifies the metadata of the page, the tmpfs file, and the VMA to which the page belongs, such that the page appears to belong to the tmpfs file and the VMA appears to be a file mapping of the tmpfs file. After the call, the calling process can access the data in this region normally, and other processes can map the tmpfs file into their address spaces via {\tt mmap}.  A {\tt deanon} call will commonly refer to a subset of a VMAs. In this case, \da\ will split VMAs as necessary.

Page tables require page-granular sharing: a fraction of a page cannot be mapped into one or more address spaces.  Nonetheless, \da\ supports calls with offsets and lengths at byte granularity.  If partial pages occur at the beginning or end of a shared range, \da\ simply copies the partial pages to newly allocated pages in the in-memory files.

When a working set is larger than available memory, {\tt deanon} may encounter swap entries when walking a page table for pages that have been swapped out. We implement two solutions for this situation.  The simple approach is to swap in the pages first to avoid the issue entirely.  Our optimized approach, called \textit{direct swap}, directly inserts a swap entry into the tmpfs file, then modifies the metadata of the swap space and the original page table entry accordingly.

Prior to de-anonymization, memory ranges will often correspond to malloc allocations from memory regions with write access enabled.  Given these regions will now map to physical memory that other processes will soon be able to access, it is important that any process exposing data with \da\ does not take any action that would modify the data.

\subsubsection{SIPC: Shared Inter-Process Communication}
\label{sec:sipc}

SIPC extends the Apache Arrow IPC implementation (v12.0.1) to use \da\ and avoid copies. SIPC, like Arrow IPC, is responsible for writing three message types to a sink: \textit{Schema}, \textit{RecordBatch}, and \textit{DictionaryBatch}. SIPC writes the schema to a sink (for us, a small in-memory file) by copying the data; schemas are usually small relative to data, so avoiding a copy would provide minimal benefit.  Records and dictionaries, in contrast, may be large, growing with the number of rows and unique string values, respectively.

SIPC creates a single tmpfs file (with the help of \da) for each column in the table, plus one more for all dictionaries (if any).  For each RecordBatch, SIPC de-anonymizes the memory backing each partial column in the batch, transferring ownership to the appropriate file.  Using different in-memory files for each column creates opportunities later to garbage collect some output columns that are not in use even while other columns of the same table are still needed.  Finer granularity would be possible (\eg{}, a new tmpfs file for every combination of batch and column, but we wish to avoid the overhead of creating too many small files.

Instead of copying the de-anonymized data to the sink, SIPC simply writes a tuple of three integers: (1) file identifier, (2) offset into file, (3) length of file range. On the read side, SIPC uses this information to make appropriate {\tt mmap} 
calls and reconstruct a view of the table without copying.  During read, SIPC also records the address ranges returned from {\tt mmap}.  Upon later writes, SIPC implements resharing by identifying output data that overlaps with the input ranges.  In this case, the output SIPC file will reference the same tmpfs files referenced by the input SIPC file; there is no need to de-anonymize a second time.

\subsubsection{Node Sandbox}
\label{sec:node-sandbox}

For our sandbox implementation, we use SOCK containers~\cite{ol-sock} with some modifications. SOCK containers are implemented in Go but are designed to run Python code. SOCK allows each container to have its own set of dependencies, but when package versions happen to be the same, installations are efficiently shared between different containers.  SOCK containers supports zygote provisioning and reusing containers for multiple invocations, but these features provide minimal value for our use case, so we disable them.

We modify SOCK to bind mount the {/dev/shm} tmpfs into every container instance to facilitate sharing (like Nightcore~\cite{Jia-21-Nightcore}).  SOCK already uses cgroups to specify memory limits, but we expose this so \zc{} can implement limit dropping to swapout specific data.  Given tmpfs memory for a file is charged to the cgroup of the process creating the file, we modify SOCK so that cgroups are retained after the process in a container exits, so outputs can be evicted from  physical memory later.  SOCK has an option to reuse the same cgroups for different containers (to avoid the cost of cgroup creation); we leave this option off so that tmpfs memory usage is never mis-charged to the incorrect node.

We also modify SOCK to provide Arrow-oriented entrypoints.  In particular, \zc{} communicates with a wrapper inside each container via {\tt sendmsg} and {\tt recvmsg} system calls over UNIX file sockets.  These allow the passing of references to SIPC Arrow data (in tmpfs) in both directions.  The wrapper is responsible for all interactions with SIPC, such that user code is only responsible for accepting and returning Arrow data.  As mentioned (\S\ref{sec:deanon}), memory that is de-anonymized should not be afterwards freed or modified, even though the memory protections will usually allow writes.  User code does not need to be aware of this constraint, however, as our generic wrapper is the last code that runs in a container, and the wrapper does not free Arrow data that has been sent via SIPC.

\subsubsection{DeCache: Deserializing Parquet}
\label{sec:decache}

We implement the DeCache in Go, leveraging existing components.  When a DAG specifies that a node should load a specific Parquet file, we run a Node Sandbox (\S\ref{sec:node-sandbox}) where the ``user code'' is actually a function we provide that uses PyArrow's {\tt parquet.read\_table} function to load the Parquet file to Arrow.  The function accepts a \texttt{read\_dictionary} argument specifying which columns should be deserialized using dictionary encoding (we construct this argument based on DAG configuration). Due to resharing, DeCache outputs may be in use, even when there are currently no nodes directly consuming the data. Tracking such references and making eviction decisions are responsibilities of the resource manager (\S\ref{sec:resource-manager}).

\subsubsection{Resource Manager}
\label{sec:resource-manager}

The Resource Manager (RM) is implemented in Go and interacts with other subsystems to gather information and perform memory-management actions.

SIPC (\S\ref{sec:sipc}) records references to tmpfs files that it has written to output, and it exposes this information to the share wrapper inside a node sandbox (\S\ref{sec:node-sandbox}), which in turn passes reference metadata back to the RM.  The RM can thus associate Arrow outputs with sets of tmpfs files.  Due to resharing and de-anonymization at column granularity, this is a many-to-many correspondence.  Visibility into these relationships allows the RM to perform reference counting on tmpfs files and garbage collect a tmpfs file backing a column of data when there are no more unrun children that will use intermediate data referencing that column.  DeCache outputs are handled differently: columns of loaded Parquet data are not immediately garbage collected upon a zero reference count because future DAGs may use the data. Data is garbage collected with a simple deletion of the relevant tmpfs file.

The RM uses \textit{limit dropping} as one of its eviction mechanisms.  Our node sandbox implementation (\S\ref{sec:node-sandbox}) allows the memory limit to be dynamically set.  Thus, the RM retains sandboxes for completed nodes and drops their limits to swapout node output as needed.  After swapping is completed, the RM raises the limit to its previous value so that the sandbox's cgroup can be charged as normal when its data is swapped back later.
\beforesect
\section{Evaluation}
\label{sec:eval}
\beforesect

\begin{table}[]
\caption{{\bf Hardware for Evaluation}. Note that the actual memory limit of each experiment is enforced by cgroup, not the RAM size. Input parquet files reside on one disk and the other is used as the swap device. SMT is disabled on CPUs.}
\begin{tabular}{c|l}
 		
CPU                 & Two Intel Xeon Silver 4314 16-core CPU     \\ \hline
RAM                 & 256GB ECC DDR4-3200 Memory \\ \hline
DISK & Two 960GB Samsung PCIe4 x4 NVMe SSD                  \\ \hline
OS                  & Ubuntu 22.04, kernel version 5.15                                     
\end{tabular}
\label{tbl:hardware}
\end{table}

We now evaluate \zc{} with a variety of workloads to answer the following: {\it how much faster can intermediate data be passed if copies are avoided} (\S\ref{sec:intermediate-data})?  {\it Does the DeCache reduce memory consumption of functions, allowing more concurrent execution} (\S\ref{sec:deserialization-cache})?
{\it How does resharing reduce memory consumption for deep DAGs} (\S\ref{sec:reshare})?  {\it Which admission and eviction techniques perform best for different scenarios} (\S\ref{sec:memory-pressure})?

Table~\ref{tbl:hardware} describes our experimental setup. The baseline we use is \zc{} with all our novel features disabled. The baseline uses regular Arrow IPC and read-side memory maps for data passing (as in Figure~\ref{fig:zero-copy}C). The baseline Resource Manager (RM) only uses admission control, data passing and reference counting, without the DeCache or eviction mechanisms (limit dropping and rollback).  For all experiments, we load input Parquet files from local storage to avoid the noisy performance often associated with cloud storage. Unless otherwise stated, the memory limit for the system is 50~GB.

\beforesect
\subsection{Intermediate Data: Copy Avoidance}
\label{sec:intermediate-data}

When a parent node produces anonymous Arrow output for children to consume, \zc{} avoids write-side copying by de-anonymizing and using the SIPC to facilitate sharing. To quantify the resulting speedup, we run a single-function DAG that that uses PyArrow to deserialize a 0.5~GB Parquet file (compressed) to about 4.0~GB of Arrow data (uncompressed), which is used for the output data. Additional memory is temporarily required during processing, such that peak memory during load is 5.8~GB.

\begin{figure}[t!]
    \centering
    \includegraphics[width=\linewidth]{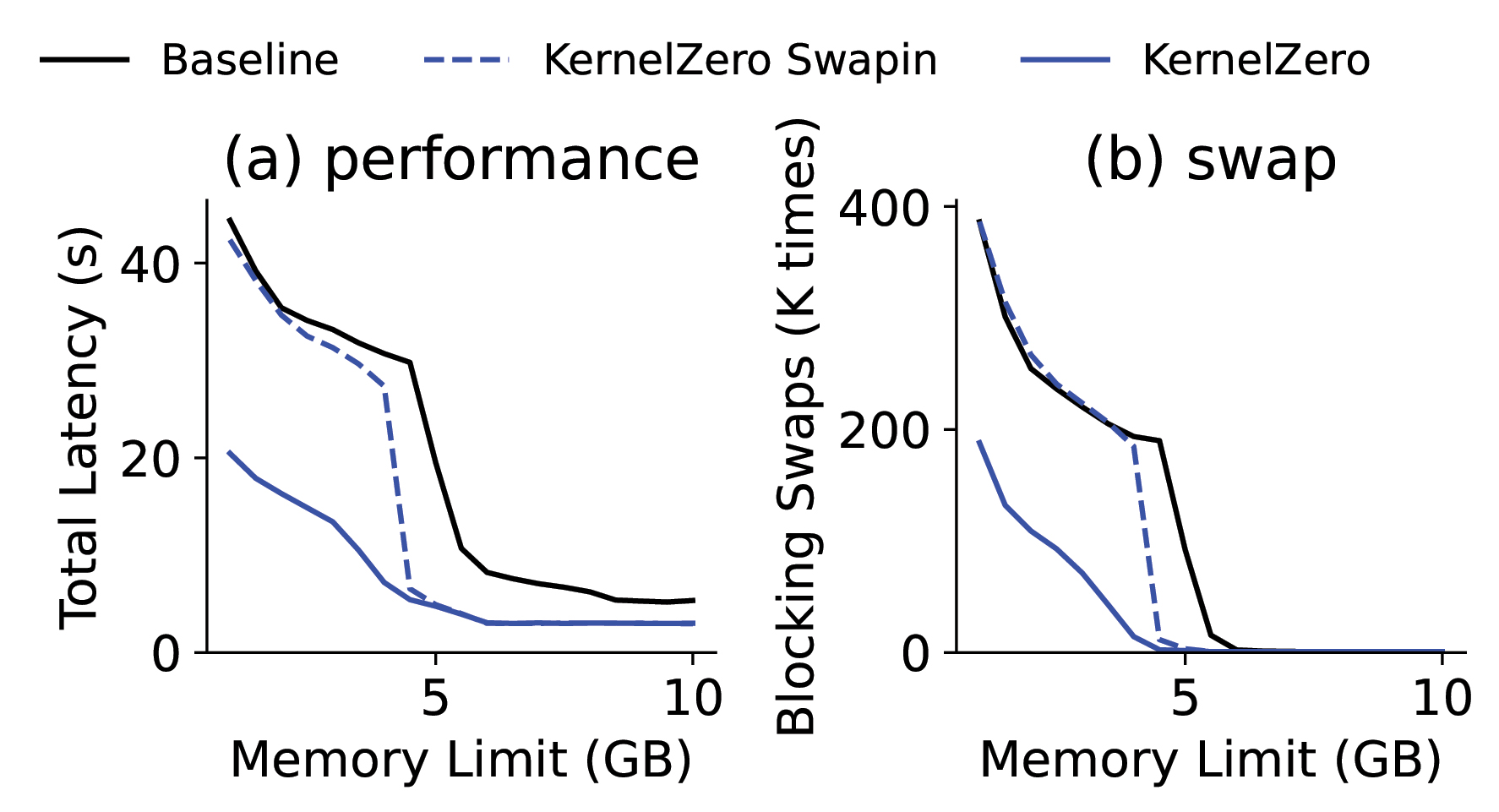}
    \vspace{-0.6cm}
    \caption{\textbf{Copy Avoidance.}  Throughput and swapping are show with and without \da\ for a single-node DAG.}
    \label{fig:eval-swap-v2}
\end{figure}

Figure~\ref{fig:eval-swap-v2}a shows the results with varying memory limits: \da{} is 1.8$\times{}$ than the baseline (based on regular Arrow IPC) when given a high memory limit (\ie{}, 10~GB) and is 2.2$\times{}$ faster for a low limit (\ie{}, 1~GB).  Although loading from a Parquet file is more work than doing a copy, the former operation is split over 24 threads for 24 columns in this table, whereas writes to an IPC file are serialized.

For high memory limits, performance gains result from copy avoidance; for low limits, Figure~\ref{fig:eval-swap-v2}b shows that the benefits result from reduced swapping.  Without \da\, copying output temporarily duplicates the data in memory, surpassing cgroup memory limits sooner. Our \textit{direct swap} technique (\da\ line) can directly copy swap entries from a page table to a tmpfs file.  Without this critical optimization,  \da\ cannot outperform the baseline under tight memory constraints.

\beforesect
\subsection{DeCache: Input Data Deduplication}
\label{sec:deserialization-cache}

\begin{figure}[t!]
    \centering
    \includegraphics[width=\linewidth]{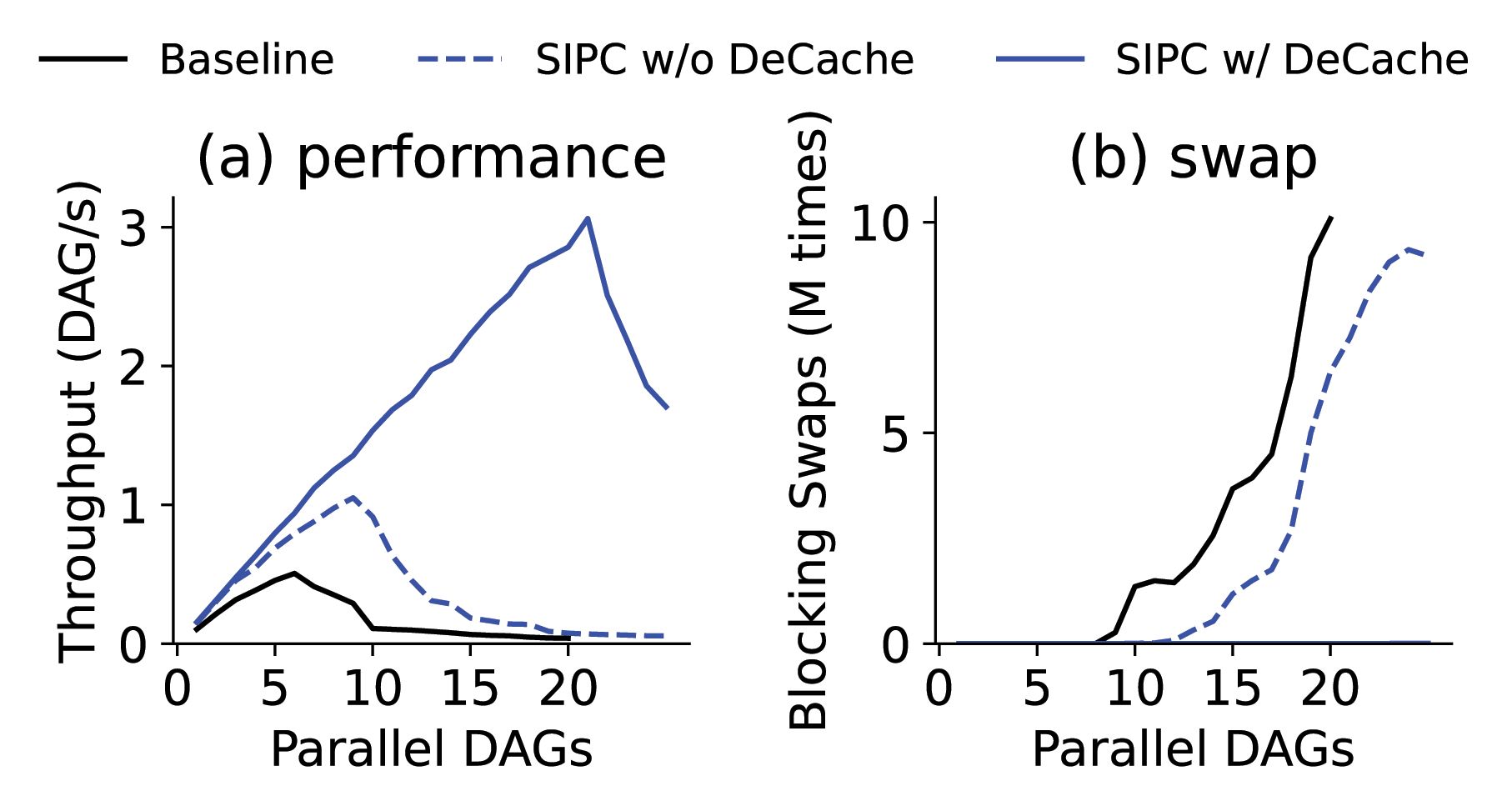}
    \vspace{-0.6cm}
    \caption{\textbf{Performance of DAGs with the same Inputs.}
        X-axis is the number of parallel executions and y-axis is the throughput in (a) and the number of foreground swap-in events in million times in (b). Baseline crashes at x$>$20 because of OOM.}
    \label{fig:parallel-dedup}
\end{figure}

Figure~\ref{fig:parallel-dedup} runs a load function loading 1.5~GB data of strings into memory, followed by a function that filters the table by rows. Multiple such DAGs (1 to 25) are sent to the worker and all the load functions load the same parquet file. There is no admission control or eviction in this experiment. SIPC w/ DeCache detects the duplicated loads to the same input across multiple DAGs, run the load only once and cache the result to be used by all the following computation functions. Thus there is no concurrent loads steps eating up the memory and the performance is significantly improved when the number of parallel DAGs are larger, up to 7.3$\times$.
\beforesect
\subsection{Resharing: Internal Copy Avoidance}
\label{sec:reshare}

When SIPC reads inputs data, it records the address ranges where it maps specific tmpfs files that were generated by upstream nodes.  During subsequent writes, SIPC is able to reshare data by reusing references to the inputs, without copying or making additional calls to \da\.  In this section, we explore the impact resharing has on performance and the size of intermediate data for a variety of operations.

\begin{figure}[t!]
    \centering
    \includegraphics[width=\linewidth]{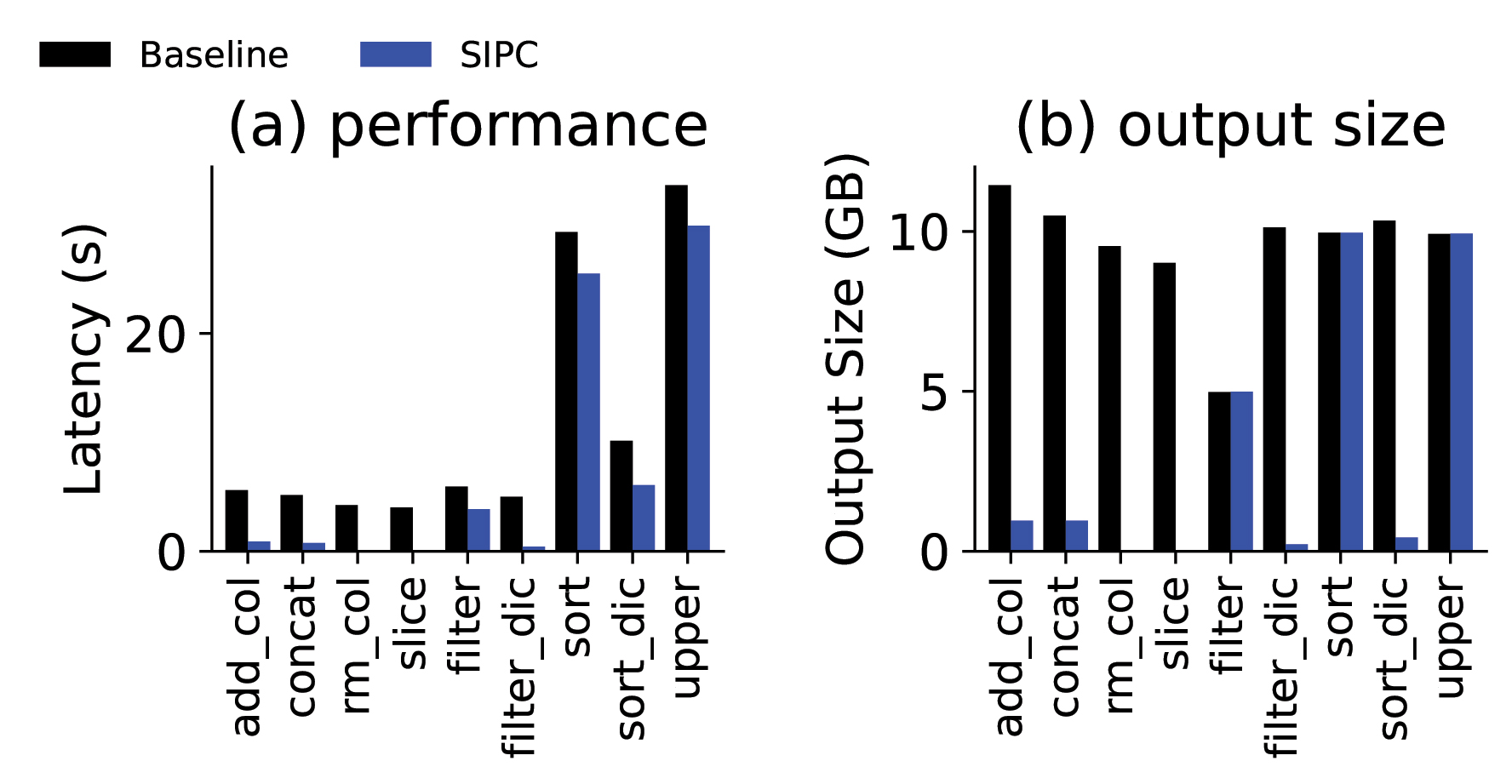}
    \vspace{-0.5cm}
    \caption{\textbf{Resharing: Time and Space Benefits}}
    \label{fig:reshare-summary}
\end{figure}

We run two-node DAGs (one load node and one compute node) for each of the operations in Figure~\ref{fig:reshare-summary}. The first 4 operations use tables of 10 1-GB integer columns as inputs, while the other 5 operates on tables of 10 1-GB string columns of 100-byte strings with no repetition. For subtractive cases (\ie{}, removing columns or slicing to obtain a consecutive subset of rows), SIPC spends almost no time and produces almost no new physical data.  For additive cases (\ie{}, adding columns or concatenating additional rows), time and space costs only apply to the additional data.

The above cases constitute coarse-grained overlap (\ie{}, similarities between inputs and outputs consist of large ranges of identical consecutive bytes).  In contrast, filtering and sorting are examples of fine-grained overlap: many of the inputs rows may appear as output rows, but with regular Arrow encodings there are not large contiguous ranges of overlap.  However, when dictionary encoding is used (``dic'' suffix on the filter and sort bars), SIPC allows us to reshare dictionaries themselves even if resharing is not possible for the buffers of numeric codes referring to dictionary entries.  We see with dictionary encoding (filter\_dic and sort\_dic), output sizes are negligible (because the dictionaries dominate the total size in our dataset), but there are no savings (relative to base line) for regular encodings (filter and sort).  The biggest performance gains are most pronounced for the dictionary encoding cases as well; though filter and sort are still somewhat better than baseline (sharing is faster than copying, even if it is slower than resharing).

In the \textit{upper} operation, all strings in a column are converted to upper case.  Unfortunately, no resharing is possible here, though it might intuitively seem possible.  Arrow string arrays have both a value buffer (containing actual characters) and an offset buffer (indicating the starting offset of each string).  If every string remained the same length after being converted to upper case, it would be possible to reshare the offset buffer, though not the value buffer; for short strings, resharing the offset buffer would be useful.  Unfortunately, Arrow arrays use UTF-8 encoding, and a few characters have different byte lengths depending on case (\eg{}, ``ß'' is 2 bytes, but the upper case equivalent is 3).  This example shows some character encodings are more reshare-friendly than others (\eg{}, the upper case optimization could be optimized to reshare offset buffers for UTF-16 strings).

\begin{figure}[t!]
    \centering
    \includegraphics[width=\linewidth]{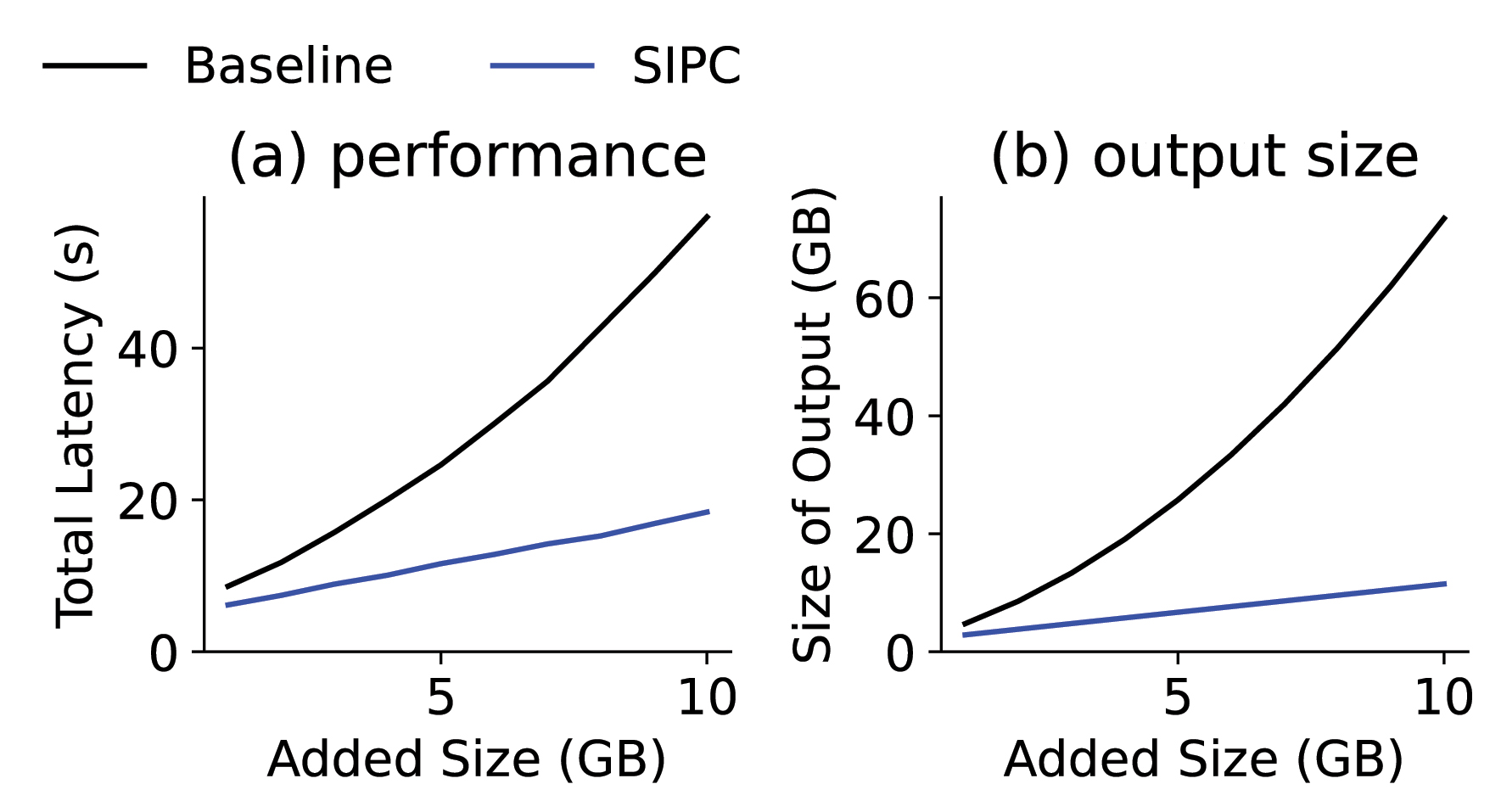}
    \vspace{-0.6cm}
    \caption{\textbf{Latency of column adding of different sizes}
          X-axis is the number of column adding function executed. Y-axis is the overall latency.}
    \label{fig:add-column}
\end{figure}

We now explore the benefits of resharing for add\_col with deeper pipelines.  We run 10 experiments with 1 to 10 column-adding functions following one load function. Each column-adding function adds a column based on computations on two randomly chosen existing columns from the previous function. The input of the workload is a 2\GB\ table of 2 1\GB columns. Each column-adding function adds a 1\GB\ column based on computations on two randomly chosen existing columns to the table and outputs the table.  

Figure~\ref{fig:add-column} shows the result.  The cumulative output size for SIPC scales linearly because each added column is only sent to tmpfs once; latency also scales linearly.  In contrast, in the baseline (without resharing), the output from any node is rewritten in every downstream node, such that intermediate sizes continue to grow throughout the pipeline.  Thus, cumulative size (and time) scales superlinearly.  Slicing exhibits the same scaling pattern as well (not shown).

\begin{figure}[t!]
    \centering
    \includegraphics[width=\linewidth]{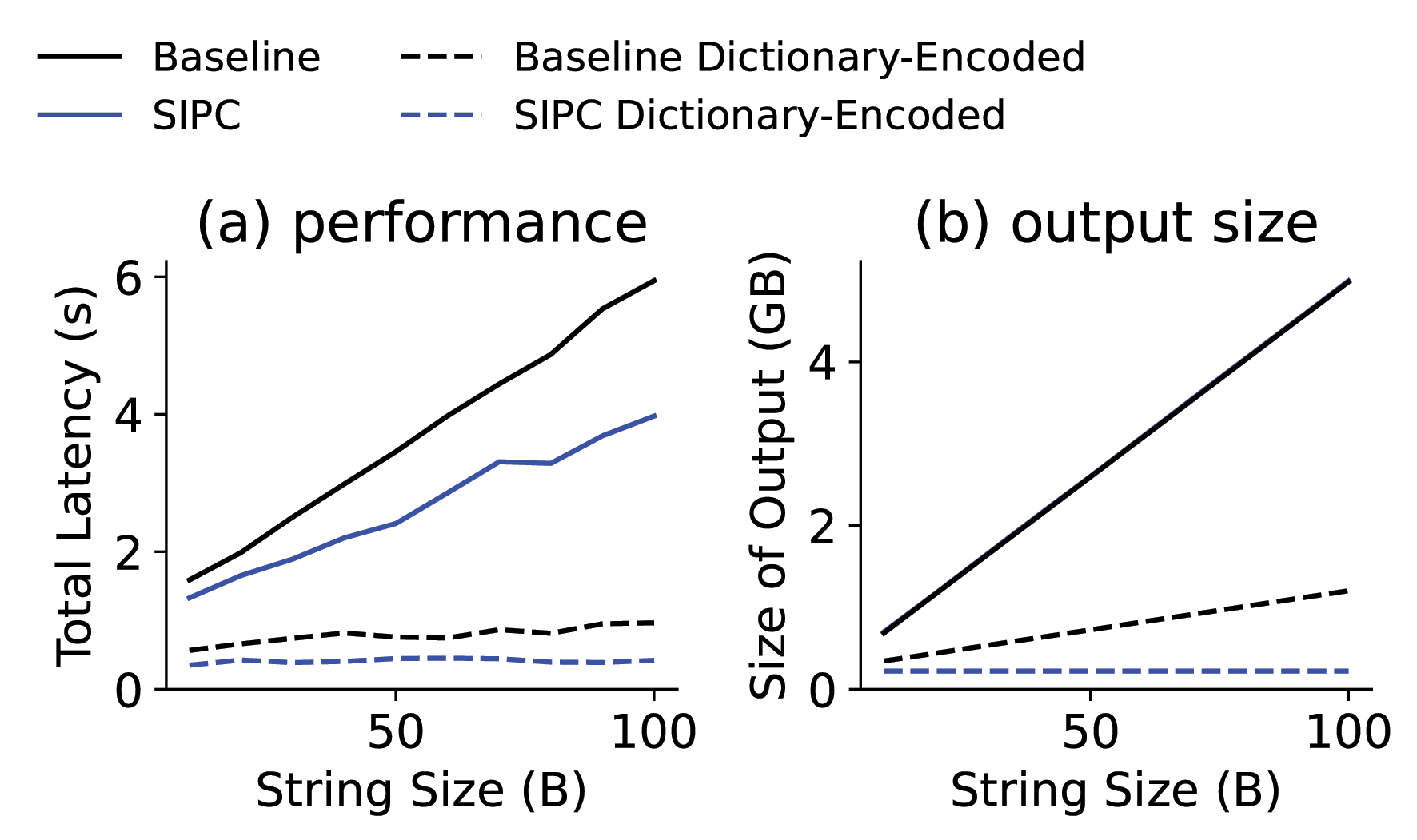}
    \vspace{-0.6cm}
    \caption{\textbf{Resharing Dictionaries (Repeats)}.  Strings are of the specified length, and each unique value appears 10 times.}
    \label{fig:filter-repeat}
\end{figure}

We now explore the interactions between dictionary encoding and resharing in more detail.  We generate a 10M-row dataset containing one column of \texttt{int32} values and 10 columns of strings of a given size.  Each unique string value occurs 10 times in a column.  After the Parquet data is loaded to Arrow (with or without dictionary encoding), a node runs a filter operation that will match half the rows, and outputs the results.  Figure~\ref{fig:filter-repeat} shows the results.  Without dictionary encoding, neither SIPC nor the baseline have the opportunity to reshare, so they produce output of identical size (SIPC does do this faster, though, because it can simply de-anonymize its Arrow data to produce the output).  The Baseline and SIPC versions both benefit significantly (in terms of time and space) from dictionary encoding because each string can be recorded once instead of 10 times.

\begin{figure}[t!]
    \centering
    \includegraphics[width=\linewidth]{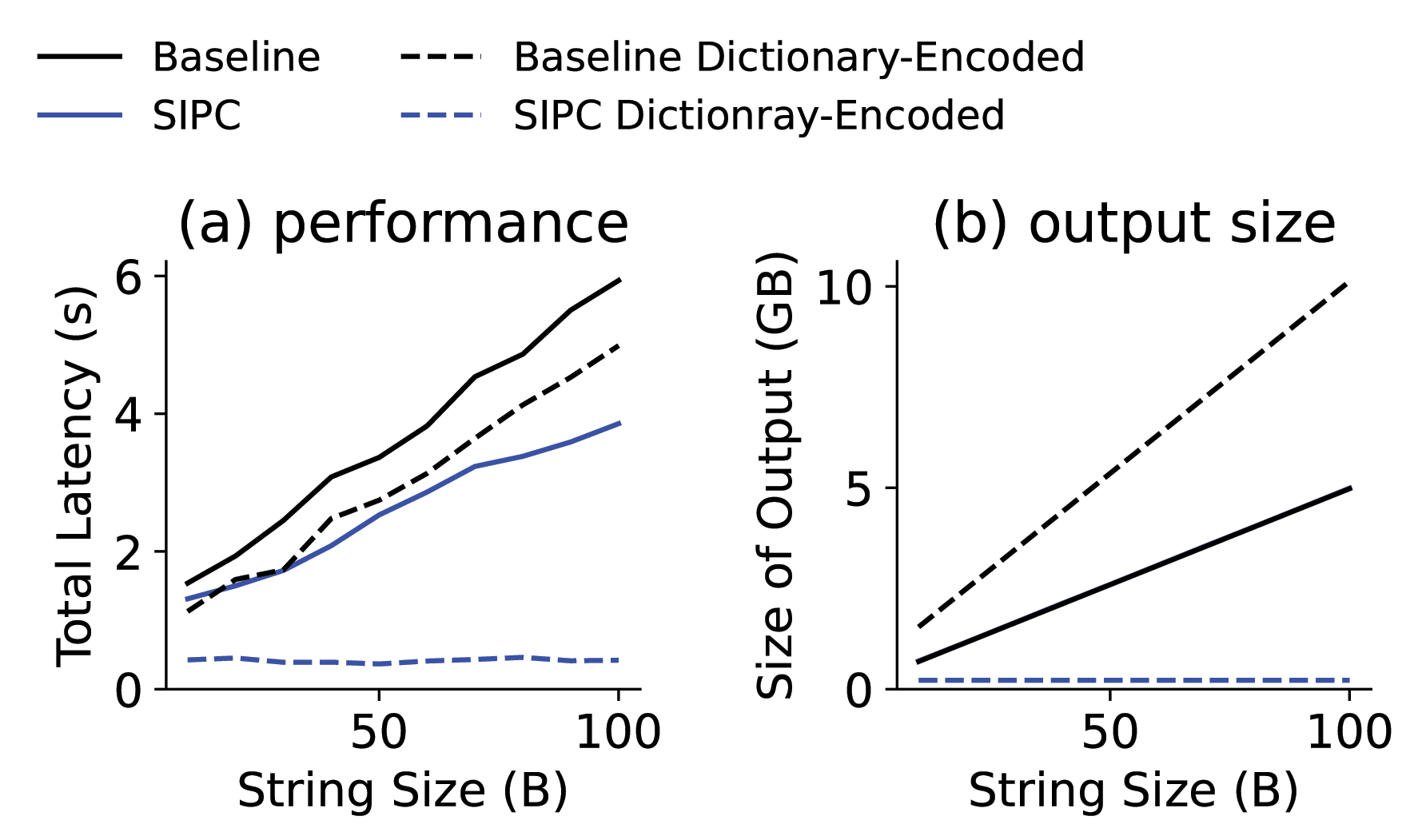}
    \vspace{-0.6cm}
    \caption{\textbf{Resharing Dictionaries (No Repeats)}.  This is the same as Figure~\ref{fig:filter-repeat}, but each string occurs only once.}
    \label{fig:filter}
\end{figure}

We perform the experiment again (Figure~\ref{fig:filter}), but this time generate the data such that each unique string occurs only once in a column.  Now, for the Baseline, the output size is actually larger with dictionary encoding: not only is there no repetition to remove, but lookup codes (into a separate table) have increased the size.  The Baseline performance still benefits from dictionary encoding, but only marginally.

In contrast, we observe that SIPC is able to reshare input dictionaries, producing intermediate outputs of negligible size extremely quickly.  We observe that \zc{}'s approach creates a compelling new reason to use dictionary encoding (besides removing redundancy in the data): supporting fine-grained resharing of values for operations such as filter and sort.
\beforesect
\subsection{Eviction Mechanisms}
\label{sec:memory-pressure}

We evaluate the mechanisms for eviction: \textit{rollback}, \textit{limit dropping}, and \textit{adaptive eviction}. We use cumulative DAGs with long chains, which have increasing memory requirements during the workload and stress eviction most.

\begin{figure}[t!]
    \centering
    \includegraphics[width=\linewidth]{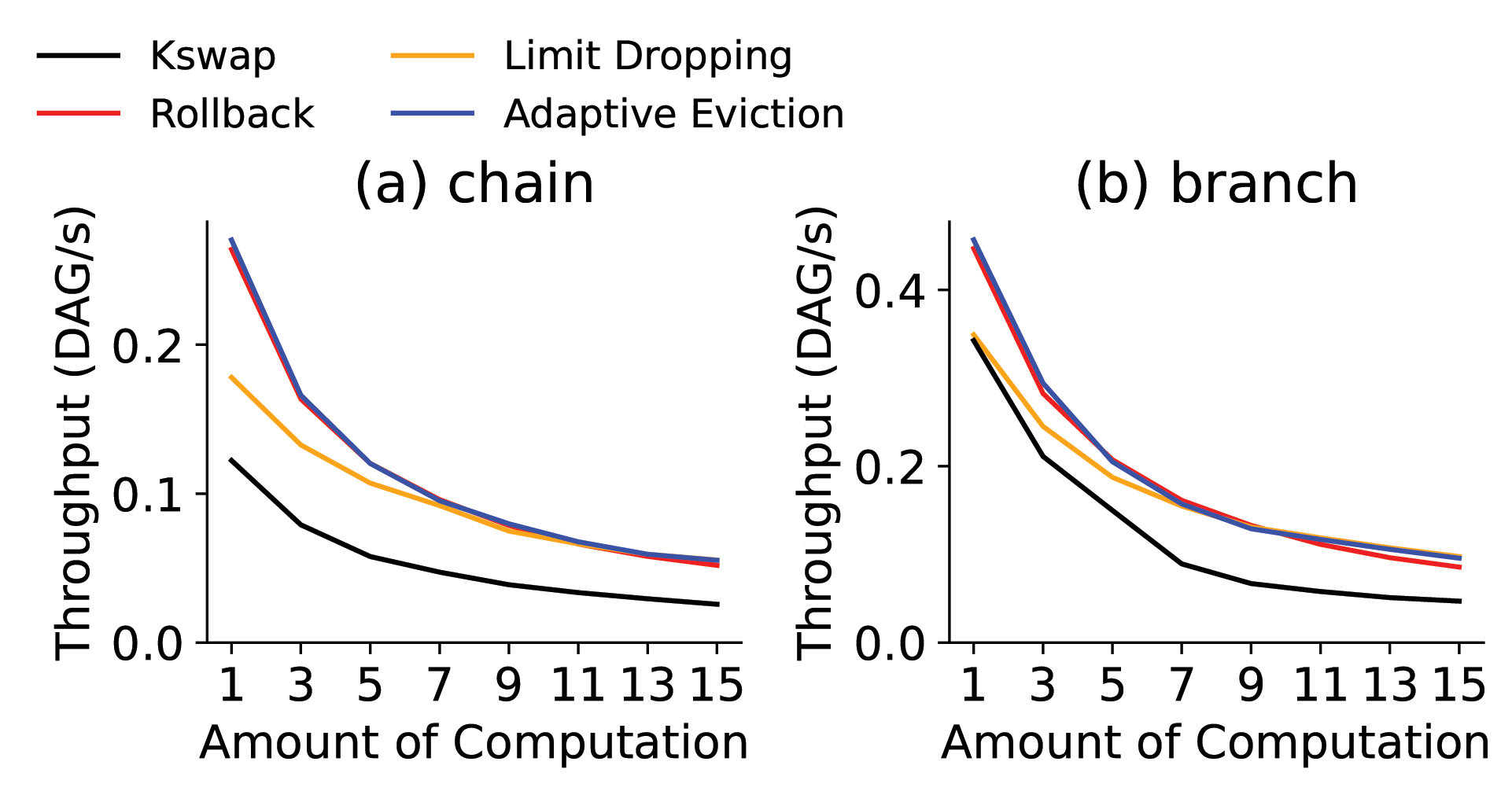}
    \vspace{-0.6cm}
    \caption{\textbf{Performance of Different Eviction Methods}
        The x-axis is the amount of computation in a single function. The y-axis is the throughput of the entire workload.}
    \label{fig:schedule-long}
\end{figure}

Figure~\ref{fig:schedule-long}a runs a workload containing 15 chains of cumulative DAGs, each containing 10 functions with the first function loading one 2~GB table and the remaining 9 appending 1~GB data to the output of the previous function based on computation of the existing data. The x-axis is the amount of computation in a single function, with unit of 1 being one simple addition of two 1~GB columns. The y-axis is the latency of finishing the entire workload. \textit{Kswap} is the baseline that only does admission control and resolves deadlocks by kernel swapping to make space for one more function to run.

The simple \textit{rollback} eviction mechanism performs much better than \textit{kswap}, with a gain of 2.0-2.2$\times$.

Comparing \textit{limit dropping} (orange line) to \textit{rollback} (red line),  when functions (which generate same amounts of outputs) runs faster for lower computation (left side of the graph), \textit{rollback} performs better, otherwise (right side of the graph) \textit{limit dropping} performs better. \textit{Adaptive eviction} (blue line) manages to choose the better eviction method and keeps the best performance for different function latencies.

Figure~\ref{fig:schedule-long}b uses different cumulative DAGs. Instead of one load function followed by add-column functions with depth 9 and fanout 1, each DAG of this workload branches out, with one load function followed by add-column functions with depth 3 and fanout 2 and a total of 15 functions in one DAG. 15 DAGs are contained in one workload. We see similar trends as Figure~\ref{fig:schedule-long}a, that \textit{rollback} performs 1.3-1.8$\times$ better than \textit{kswap}, and that \textit{adaptive eviction} performing the best under various conditions.

\beforesect
\section{Related Work}
\label{sec:related}
\beforesect

\textbf{Data Pipelines.} Many classic data-processing platforms (\eg{}, MapReduce~\cite{Dean-04-MapReduce,Shvachko-10-Hadoop} and Spark~\cite{Zaharia-12-Spark}) impose both a computational model and the intermediate data format.  Restricted environments are conducive to automatic optimization, but the need for flexibility has created use cases for other data pipeline orchestration tools, including 
Apache Airflow~\cite{Apache-Airflow}, Luigi~\cite{Luigi}, and Metaflow~\cite{Tagliabue2023ReasonableSM}.  These tools allow developers to use arbitrary approaches for data passing, at the cost of data duplication and overall inefficient I/O.

\textbf{Storage-Based Communication.} Many data processing platforms leverage distributed file systems (\eg, HDFS) or cloud storage (\eg{}, S3) for intermediate data, but experience has shown that it is difficult to achieve fast and reliable performance this way~\cite{Perron-20-Starling}.  SONIC~\cite{Mahgoub-21-SONIC} and SAND~\cite{Akkus-21-SAND} use a hybrid method of remote storage and local file system based on online profiling and function placement.  Fortunately, hardware improvements are catching up to workload sizes~\cite{Renen-24-Redshift} such that most workloads can run on a single machine and take advantage of faster mediums~\cite{McSherry2015ScalabilityBA,jordandata,10.1145/3460231.3474604}.

\textbf{Zero-Copy I/O Stacks.} There have been many efforts to avoid copies in network and storage stacks. PASTE~\cite{Honda-18-PASTE} and Fastmove~\cite{Su-23-Fastmove} use DMA to avoid copying. Arrakis~\cite{Peter-14-Arrakis} redesigns the kernel as a control plane so that I/O data may reside solely in user space. zIO~\cite{Stamler-22-zIO} deduplicates buffers for I/O transparently, with a key observation that applications often modify only small parts of the data.

\textbf{Zero-Copy Inter-Process Communication (IPC).} Manipulating page tables is a well-known approach to avoid copying.  Fbufs~\cite{Druschel-93-Fbufs} use this approach for IPC, but programs must identify what data to share from the beginning.  In contrast, \zc{} can de-anonymize data generated by arbitrary libraries after the fact. IO-Lite~\cite{Pai-99-IOLite} is an extention of Fbufs that adds a \textit{buffer aggregate} abstraction.  Like resharing, buffer aggregation allows data certain transformations (\eg{}, slicing and concatenation) to reference portions of original data.  \zc{} applies resharing in the context of Arrow, where additional techniques are possible (\eg{}, dictionary sharing and IPC inspection).  Nightcore~\cite{Jia-21-Nightcore}, like \zc{}, mounts a tmpfs into multiple containers to facilitate communication via shared memory.  Nightcore focuses on microservices and low-latency RPC-based bi-directional communication; \zc{} is optimized for DAG workloads where multiple children consume the large outputs of multiple parents.

\textbf{Linux Kernel Mechanisms.} The Linux kernel has several existing and proposed features related to our work.  KSM~\cite{KSM} scans physical memory and modifies page tables to deduplicate redundant pages it discovers.  DeCache avoids duplication from the start, and thus avoids expensive scans and short-term duplication.  \zc{}'s limit dropping approach was possible due to Linux's integration between cgroups and swap, a feature demonstrated by Senpai~\cite{meta-memory-offloading}.  Senpai's developers eventually abandoned limit dropping because the system could not respond quickly enough to surging memory needs.  \zc{} only uses limit dropping to swapout intermediate data produced by completed processes, so this approach better suites our use case.  A new  \textit{process\_vm\_mmap}~\cite{process-vm-mmap} API has been proposed for Linux that would allow different processes to share VMAs (including anonymous ones).  A \zc{}-like system could be built around such a primitive, though being able to directly access another address space would naturally be a privileged operation (a file-oriented approach such as de-anonymization supports better control over visibility).  Another proposed API (\textit{msharefs}~\cite{msharefs}) would allow different processes to share not only physical pages, but page table entries.  We expect that SIPC could be slightly faster using such a feature.
\beforesect
\section{Conclusion}
\label{sec:conc}
\beforesect

The term ``zero-copy'', as commonly used in the data ecosystem, does not actually mean no copying.  In this work, we consider often-overlooked occurrences of copying and duplication in data workloads run today on Bauplan, and pursue an ideal ``true zero copy'' mechanism for in-memory DAG execution. We introduced \zc{}, a new system that avoids write-side copies (from nodes) and input-to-output copies (within nodes). \zc{} further avoids duplication between different DAGs using the same inputs. Although some copying and duplication remains unavoidable, our preliminary benchmarks show that \zc{}'s may improve throughput significantly in a variety of data scenarios.

\clearpage
{
   \small
   \bibliographystyle{plain}
   \bibliography{all-defs,all,urls,all-confs,other}
}

\clearpage
\end{spacing}
\end{document}